%% file: main.tex
\documentclass[sigconf,nonacm]{acmart}
\input{Preamble/packages}

\input{Preamble/definitions}
\input{Preamble/meta}
\input{Preamble/authors}

\begin{document}

\title[Retrieval-in-the-Chain: Bootstrapping Large Language Models for Generative Retrieval]{Retrieval-in-the-Chain: Bootstrapping Large Language Models for Generative Retrieval}

\begin{abstract}
Generative retrieval (GR) is an emerging paradigm that leverages large language models (LLMs) to autoregressively generate document identifiers (docids) relevant to a given query. 
Prior works have focused on leveraging the generative capabilities of LLMs to improve GR, while overlooking that their reasoning capabilities could likewise help. 
This raises a key question: Can explicit reasoning benefit GR? 
To investigate, we first conduct a preliminary study where an LLM is prompted to generate free-form chain-of-thought (CoT) reasoning before performing constrained docid decoding.
Although this method outperforms standard GR, the generated reasoning tends to be verbose and poorly aligned with the docid space. 
These limitations motivate the development of a reasoning mechanism better tailored to GR.

Therefore, we propose Reason-for-Retrieval (R4R), a reasoning-augmented framework for GR that converts free-form CoT reasoning into a compact, structured format, and iteratively refines the reasoning during the retrieval process. 
R4R augments an existing GR method by leveraging a reasoning-capable LLM that has been instruction-tuned for GR. 
At inference time, R4R first uses the LLM to generate an initial structured reasoning; then the same LLM alternates between (i) constrained decoding with the chosen GR method to produce candidate docids and (ii) updating the reasoning based on retrieval results to improve the next round. 
R4R does not require additional models or training, and instead a single LLM serves as both the reasoning generator and the retriever. 
Extensive experiments on Natural Questions, MS MARCO, and a real-world item-search benchmark validate the effectiveness of R4R.

\end{abstract}

\maketitle

\input{Sections/1-intro}

\input{Sections/3-exploration_challenges}

\input{Sections/4-method}

\input{Sections/5-experimental_setup}
\input{Sections/6-experiments}

\input{Sections/2-related_works}

\input{Sections/7-conclusion}

\bibliographystyle{ACM-Reference-Format}
\balance
\bibliography{references}
\input{Sections/appendix}

\end{document}

%% file: Preamble/packages.tex
\usepackage{color}
\usepackage{bbm}
\usepackage{multirow}

\usepackage[inline]{enumitem}
\usepackage{graphicx}
\usepackage{subcaption}
\usepackage{sistyle}
\usepackage{booktabs}
\usepackage{caption}
\SIthousandsep{,}
\usepackage{makecell}
\usepackage{amsmath}
\usepackage[english]{babel}
\usepackage{hyperref}
\usepackage{array} 
\usepackage{graphicx}
\usepackage{tikz}
\usepackage{wrapfig}
\usepackage{acronym}
\usepackage{graphicx}  %插入图片的宏包
\usepackage{float}  %设置图片浮动位置的宏包
\usepackage{subcaption}  %插入多图时用子图显示的宏包
\usepackage{fontawesome}
\usepackage[most]{tcolorbox}

\usepackage[ruled,vlined,linesnumbered]{algorithm2e}

\usepackage{amssymb} % 提供 \triangleright
\SetKwComment{rtri}{$\triangleright$ }{}
\SetKwComment{rbtri}{$\blacktriangleright$ }{}

%% file: Preamble/definitions.tex
\newcommand{\heading}[1]{\smallskip\noindent\textbf{#1.}}

\AtBeginDocument{%
  \providecommand\BibTeX{{%
    \normalfont B\kern-0.5em{\scshape i\kern-0.25em b}\kern-0.8em\TeX}}}

\makeatletter
\g@addto@macro\normalsize{%
  \abovedisplayskip 3pt plus1pt %minus1pt%
  \belowdisplayskip 3pt plus1pt
  \abovedisplayshortskip  0pt plus1pt%
  \belowdisplayshortskip  0pt plus1pt% minus1pt%
}
\makeatother

\acrodef{CV}{computer vision}
\acrodef{IR}{information retrieval}
\acrodef{LLM}{large language model}
\acrodef{MDP}{Markov decision process}
\acrodef{NLP}{natural language processing}
\acrodef{NRM}{neural ranking model}
\acrodef{RL}{reinforcement learning}
\acrodef{RL-MARA}{Multi-grAnular Ranking Attack}
\acrodef{MoE}{mixture-of-experts}

%% file: Preamble/meta.tex
\copyrightyear{2025}
\acmYear{2025}
\setcopyright{rightsretained}
\acmConference[The Web Conference '26]{The 2026 ACM Web Conference}{April 13--17, 2026}{Dubai, UAE}
\acmBooktitle{The 2026 ACM Web Conference, April 13--17, 2026, Dubai, UAE}
\acmDOI{}
\acmISBN{}

\makeatother
 
\ccsdesc[500]{Information systems~Retrieval models and ranking}

\keywords{Generative retrieval, Chain-of-thought, Iterative refinement}

%% file: Preamble/authors.tex
\author{Yingchen Zhang}
\orcid{0009-0003-3979-6069}
\affiliation{
 \institution{State Key Laboratory of AI Safety, Institute of Computing Technology, Chinese Academy of Sciences}
 \institution{University of Chinese Academy of Sciences}
 \city{Beijing}
 \country{China}
}
\email{zhangyingchen23s@ict.ac.cn}

\author{Ruqing Zhang}
\orcid{0000-0003-4294-2541}
\authornote{Jiafeng Guo and Ruqing Zhang are the corresponding authors.}
\affiliation{
 \institution{State Key Laboratory of AI Safety, Institute of Computing Technology, Chinese Academy of Sciences}
 \institution{University of Chinese Academy of Sciences}
 \city{Beijing}
 \country{China}
}
\email{zhangruqin@ict.ac.cn}

\author{Jiafeng Guo}
\orcid{0000-0002-9509-8674}
\authornotemark[1]
\affiliation{
 \institution{State Key Laboratory of AI Safety, Institute of Computing Technology, Chinese Academy of Sciences}
 \institution{University of Chinese Academy of Sciences}
 \city{Beijing}
 \country{China}
}
\email{guojiafeng@ict.ac.cn}

\author{Wenjun Peng}
\orcid{0000-0003-2392-4946}
\affiliation{
 \institution{Researcher}
 \city{Hangzhou}
 \country{China}
}
\email{pengwj@mail.ustc.edu.cn}

\author{Sen Li}
\orcid{0000-0002-2124-953X}
\affiliation{
 \institution{Researcher}
 \city{Hangzhou}
 \country{China}
}
\email{lisen.lisen@alibaba-inc.com}

\author{Fuyu Lv}
\orcid{0000-0001-5918-093X}
\affiliation{
 \institution{Researcher}
 \city{Hangzhou}
 \country{China}
}
\email{fuyu.lfy@alibaba-inc.com}

\renewcommand{\shortauthors}{Yingchen Zhang et al.}

%% file: Sections/1-intro.tex
\section{Introduction}
\label{sec:intro}

Generative retrieval (GR) \cite{dsi,seal,gere,genre,nci} has emerged as a new paradigm in information retrieval (IR), distinct from traditional dense retrievers \cite{dpr,colbert} based on dual encoders. 
Specifically, GR employs an autoregressive generative model to implicitly encode the corpus within its parameters during training. 
During inference, it generates document identifiers (docids) directly from the query in an end-to-end manner, producing valid candidates under prefix constraints \cite{dsi}.

\begin{figure}[t]
    \centering
    \includegraphics[width=\linewidth]{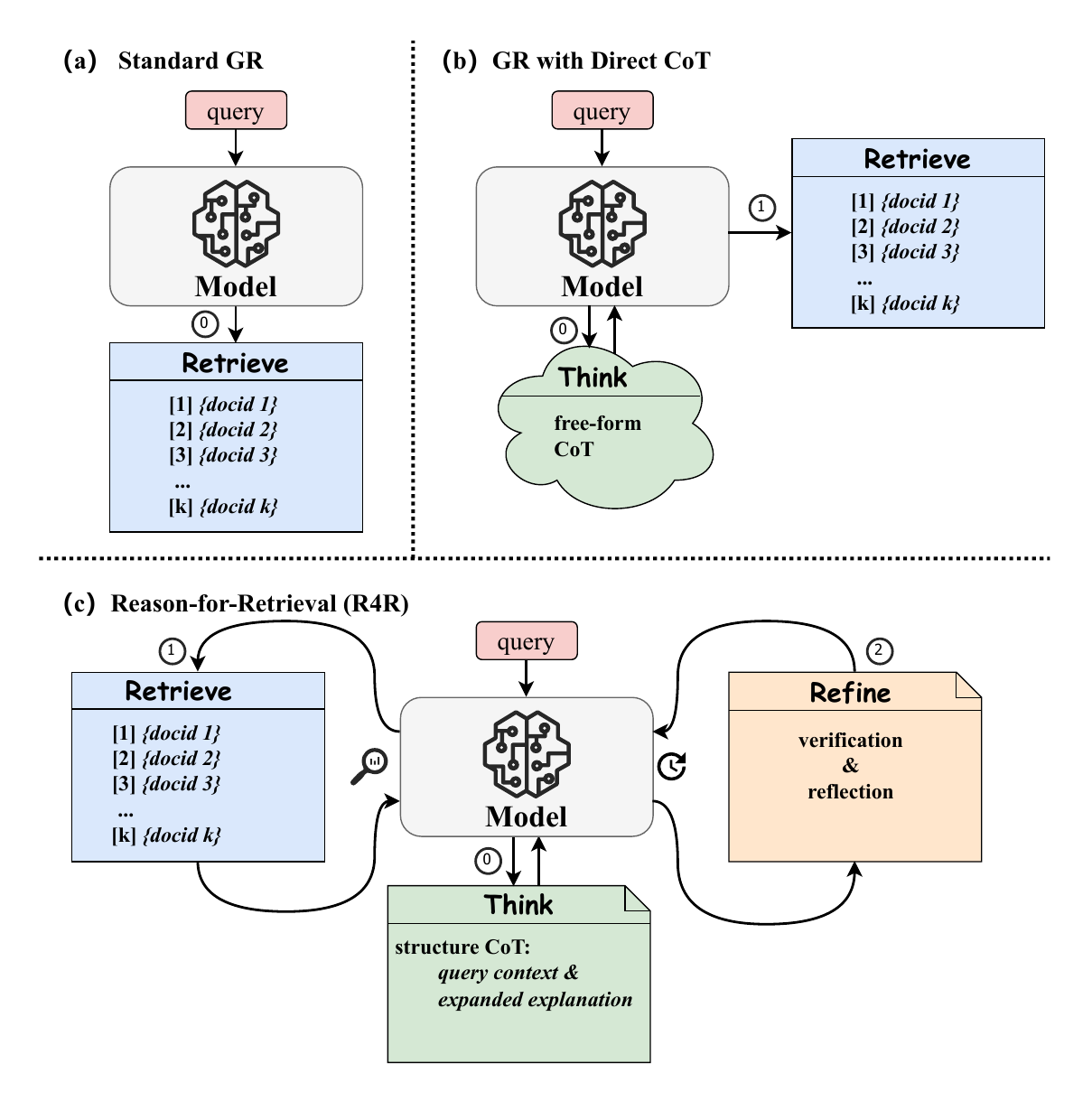}
    \caption{Comparison of (a) standard GR, (b) GR + Direct CoT, and (c) our proposed R4R. R4R compresses and structured reasoning, forming an iterative improvement pipeline}.
    \label{fig:intro}
    \vspace{-4mm}
\end{figure}

Through pretraining on large-scale text corpora and alignment fine-tuning to follow human instructions, large language models (LLMs), such as GPT \cite{gpt} and Qwen \cite{qwen2.5}, have demonstrated exceptional performance in language understanding, generation and reasoning \cite{gpt3,deepseek-r1}. 
Consequently, there is growing interest in leveraging the language understanding and generative capabilities of LLMs within GR. 
Indeed, GR has achieved performance comparable to, or even surpassing, dense retrieval on tasks such as image, code and book retrieval \cite{grmm,codedsi,grbook}.

Recent studies show that explicit step-by-step reasoning, such as Chain-of-Thought (CoT) prompting \cite{cot}, significantly enhances LLM performance in natural language processing (NLP) tasks, like question answering, mathematical reasoning, and commonsense reasoning \cite{cot,self-consistency-cot,llm-reasoner}. 
Rather than predicting answers directly, prompting the model to ``think first, then decide, and optionally reflect'' often leads to more accurate outcomes. 
However, the objectives of GR differ from conventional NLP tasks: GR models need to comprehend user intent, perform global comparisons across the corpus, and generate relevant docids, all while bridging the semantic gap between queries and their corresponding docids.   
Consequently, how to effectively harness LLMs' reasoning capabilities to enhance GR remains a challenging and largely unexplored research problem.

To this end, we focus on the following questions: 
\begin{enumerate}[label=\textbf{(RQ\arabic*)},nosep,leftmargin=*]
\item What are the effects of straightforward, unstructured CoT prompting on GR?
\item How can we design a novel CoT prompting framework tailored for GR while preserving its core reasoning benefits? 
\end{enumerate}

To answer \textbf{RQ1}, we evaluate a straightforward ``Direct CoT'' strategy, illustrated in Figure~\ref{fig:intro}(b). 
During inference, the LLM is first prompted to generate free-form CoT reasoning about the query. 
This reasoning is then concatenated with the original query, and standard constrained decoding is applied to produce docids. 
To retain the LLM's generative capabilities even after being trained for retrieval, we adopt two design measures: 
(i) we use textual rather than numeric docids, ensuring the model's output remains within the natural language space; and 
(ii) unlike standard GR training \cite{dsi,seal} where only the query is typically used as input, we explicitly prepend a retrieval-specific instruction to the query, signaling to the LLM that it is operating in retrieval mode.
Experimental results under this setup confirm that Direct CoT outperforms the standard GR baseline.  
However, we also observe that free-form CoT tends to be verbose, which not only increases inference latency but also introduces noise into the retrieval.

To answer \textbf{RQ2}, we introduce Reason-for-Retrieval (R4R), a iterative reasoning-augmented framework for GR, as illustrated in Figure~\ref{fig:intro}(c). 
Its core idea is to reformulate the model's free-form reasoning into a structured and compact format tailored for retrieval, which is then progressively refined. 
R4R leverages a reasoning-capable LLM and a GR method. 
Firstly, the LLM is instruction-tuned using the GR method to align its retrieval behavior with specific prompts
Then, at inference time, given an input query, R4R begins with a \emph{Think} step that generates formatted initial reasoning, comprising a concise \emph{query context} and an \emph{expanded explanation} of the query intent. The model then alternates between two subsequent steps: 
(i) \emph{Retrieve}: standard constrained decoding is performed using both the query and the current query context to produce candidate docids; and
(ii) \emph{Refine}: the model reflects on the error by analyzing the output irrelevant docid, and updates both the query context and the expanded explanation for the next iteration. 
The iterative loop terminates when a predefined round limit is reached or when the top candidates are confirmed to be relevant.

R4R could be seamlessly integrated into existing GR methods using text-based docids with minimal engineering overhead. 
Empirical evaluations conducted on Natural Questions (NQ) \cite{nq} and MS MARCO \cite{msmarco} demonstrate consistent improvements when R4R is integrated with four mainstream GR methods. 
The results confirm that R4R enhances retrieval performance without introducing substantial decoding overhead. 
Further experiments on a real-world Taobao item-search benchmark validate these gains, underscoring R4R's practical applicability in industrial scenarios.

%% file: Sections/3-exploration_challenges.tex
\vspace{-2mm}
\section{Exploration and challenges}
\label{sec:exploration}

In this section, we first introduce the standard GR framework, covering its training and inference procedures as well as docid construction. 
We then describe how GR is adapted to operate within the reasoning process. 
Finally, we present experimental results to investigate whether Direct CoT can enhance GR performance.

\subsection{Standard GR}
\label{subsec:standard_gr}
\heading{Training}
Standard GR training involves two basic operations: 
(i) indexing $\mathcal{L}_{\text{GR}}^{indexing}$, which memorizes the entire corpus $\mathcal{D}$ by associating each document $d$ with its corresponding identifier $docid(d)$; 
and (ii) retrieval $\mathcal{L}_{\text{GR}}^{retrieval}$, which uses the indexed corpus information to predict relevant docids for a given query $q$, defined as: 
\begin{subequations}
\label{eq:standard_training}
\begin{align}
\mathcal{L}_{\text{GR}}^{indexing}
&= - \sum_{d\in\mathcal{D}} \sum_{i=1}^{L}
\log p_{\mathcal{M}}\!\left(docid(d)_i \,\middle|\, docid(d)_{<i},\, d\right), \\
\mathcal{L}_{\text{GR}}^{retrieval}
&= - \sum_{(q,\,d)\in\mathcal{D}} \sum_{i=1}^{L}
\log p_{\mathcal{M}}\!\left(docid(d)_i \,\middle|\, docid(d)_{<i},\, q\right),
\end{align}
\end{subequations}
where $L$ denotes the length of docid and $\mathcal{M}$ denotes the GR model. 
The overall training objective $\mathcal{L}_{\text{GR}}$ is usually defined as the sum of indexing objective $\mathcal{L}_{\text{GR}}^{indexing}$ and retrieval objective $\mathcal{L}_{\text{GR}}^{retrieval}$, i.e., $\mathcal{L}_{\text{GR}} = \mathcal{L}_{\text{GR}}^{indexing}+\mathcal{L}_{\text{GR}}^{retrieval}$. 
While effective for retrieval, this approach can degrade the model's general text-generation ability, thereby hindering reasoning generation during inference. 

\heading{Inference}
During standard GR inference, the model takes a query as input and performs constrained beam search to generate a set of candidate docids (different GR methods use different constraint strategies, e.g., prefix-trie \cite{dsi,nci} and FM-index \cite{seal,minder}). Formally,
\begin{equation}
docid[1{:} k] = \mathcal{M}(q;\mathrm{cons}),
\label{eq:standard_inference}
\end{equation}
where $k$ denotes the beam width (also the number of generated candidates), and $\mathrm{cons}$ denotes the constraint strategy.

\heading{Docid construction} 
Existing works on docid design typically include two types: 
(i) \emph{numeric docids:} assign each document a unique integer \cite{dsi} or a semantic codebook \cite{dsi,nci} string as its identifier; and 
(ii) \emph{textual docids:} use titles \cite{seal,corpus-brain}, n-grams \cite{tsgen}, or pseudo-queries \cite{minder} as identifiers.

\subsection{Adapting GR for reasoning}
\label{subsec:adapted_gr}
\heading{Adapted training}
To preserve the model's generative versatility, we adopt an instruction-tuning strategy for both indexing and retrieval training. 
We prepend a task-specific instruction prompt to the input, for example, using $P_i$ for indexing operation and $P_r$ for retrieval operation (see Appendix~\ref{app:prompt} for templates), i.e., 
\begin{subequations}
\label{eq:instruction_training}
\begin{align}
\mathcal{L}_{\text{GR}_{ins}}^{indexing}
&= - \sum_{d\in\mathcal{D}} \sum_{i=1}^{L}
\log p_{\mathcal{M}}\!\left(docid(d)_i \,\middle|\, docid(d)_{<i},\, d,\,P_i\right), \\
\mathcal{L}_{\text{GR}_{ins}}^{retrieal}
&= - \sum_{(q,\,d)\in\mathcal{D}} \sum_{i=1}^{L}
\log p_{\mathcal{M}}\!\left(docid(d)_i \,\middle|\, docid(d)_{<i},\, q,\,P_r\right).
\end{align}
\end{subequations}

\heading{Adapted inference}
At inference time, the retrieval prompt $P_r$ is prepended to the query before constrained decoding, explicitly signaling the model to operate in retrieval mode.
Formally,
\begin{equation}
docid[1{:} k] = \mathcal{M}\big(P_r\,\|\ \,q \,\,\|\,\text{rw}(q);\;\mathrm{cons}\big),
\label{eq:instruction_inference}
\end{equation}
where $\text{rw}(q)$ represents an optional rewrite of query $q$.
This setup maximizes the preservation of the model’s native generative capabilities, while performing accurate and constrained retrieval.

\heading{Adapted docid construction} 
To adapt GR for reasoning, we employ textual docids rather than numeric docids for two main reasons:
(i) to maintain both reasoning outputs and docids within the same token space, that of natural language, thereby enabling the reasoning process to effectively guide docid generation; and
(ii) to prevent the LLM from losing its general text-generation capability after retrieval-oriented training, which could occur if it were restricted to producing only numeric docids.

To ensure broad compatibility, we do not impose a specific textual docid format but instead align with any chosen GR method. 
In cases where a GR method does not prescribe a particular textual docid scheme (or only prescribe it for a specific dataset), we define default docids for R4R as follows:
(i) build a document hierarchy using residual quantization (RQ) \cite{rq};
(ii) assign a unique word (obtained via keyword extraction) to each node in the hierarchy as its level-specific identifier; and
(iii) form the final docid by concatenating the words along the path from the root node to the target document.

\subsection{Direct CoT for GR}

\label{subsec:direct_cot}
\heading{Direct CoT: inference-only} 
After training GR models under the above adapted GR setup, a straightforward way to integrate reasoning into GR is to directly prompt the model to generate reasoning for the query and use it to guide retrieval. 
Specifically, we perform a two-step inference process:
(i) First, we prompt the model to generate free-form CoT reasoning using unconstrained decoding prompt $P_d$, an approach we refer to as Direct CoT (see Appendix~\ref{app:prompt} for prompt). 
(ii) Subsequently,  candidate docids are produced through constrained decoding, using both the original query and the generated reasoning as context. 
Formally, 
\begin{subequations}
\label{eq:direct_cot}
\begin{align}
&DC = \mathcal{M}(P_d\,\| \,q),\\
&docid[1{:} k] = \mathcal{M}(P_r\,\|\ \,q\,\|\ \;DC;\;\mathrm{cons}), 
\end{align}
\end{subequations}
where DC denotes the generated Direct CoT, playing a role similar to $\text{rw}(q)$ in Eq.~\ref{eq:instruction_inference}.

\heading{Experiments}
We conduct GR experiments on the NQ dataset under three settings: 
(i) standard GR: trained as in Eq.~\ref{eq:standard_training} and inferred as in Eq.~\ref{eq:standard_inference}; 
(ii) standard GR with CoT: trained as in Eq.~\ref{eq:standard_training} and inferred as in Eq.~\ref{eq:direct_cot}; and 
(iii) adapted GR with CoT: trained as in Eq.~\ref{eq:instruction_training} and inferred as in  Eq.~\ref{eq:direct_cot}. 
All experiments use R4R’s default docid and a DSI-style prefix-trie constraint \cite{dsi}.

\input{tables/exploration}

As shown in Table \ref{tab:exploration}, we observe that: 
(i) standard GR with Direct CoT performs far worse than standard GR, because standard GR trained with direct mapping largely loses its normal generative ability and thus fails to produce useful reasoning to help retrieval; 
(ii) adapted GR (instruction-tuned) with Direct CoT outperforms standard GR, demonstrating that GR can indeed benefit from explicit reasoning; and 
(iii) Direct CoT introduces substantial latency compared with standard GR, indicating that free-form CoT incurs significant decoding overhead. 

\heading{Discussion}
These findings answer \textbf{RQ1}: while Direct CoT can modestly improve GR, it often yields verbose rationales. These greatly increase decoding costs and may introduce noise that hurts retrieval. 
The core issue is a mismatch: GR is trained on short and concise queries, while long free-form CoT introduces extraneous reasoning that drowns out key retrieval signals. 
Consequently, the model fails to leverage such reasoning.  
To address this, we redesign a structured CoT tailored for GR. It keeps reasoning concise to cut latency and aligns intermediate signals with the docid space.

%% file: tables/exploration.tex
\begin{table}[h]
\centering
\caption{Exploration of GR with Direct CoT.}
\label{tab:exploration}
\resizebox{\columnwidth}{!}{
\begin{tabular}{lccccc}
\toprule
 & Hits@1 & Hits@5 & Hits@20 & MRR@10 & Latency \\
\midrule
Standard GR           & 45.8 & \textbf{59.6} & 75.3 & 56.3 & \textbf{3.2}  \\
Standard GR + CoT     & 12.8 & 18.2 & 21.5 & 15.1 & 72.8 \\
Adapted GR + CoT  & \textbf{46.0} & \textbf{59.6} & \textbf{76.0} & \textbf{57.5} & 63.9 \\
\bottomrule
\end{tabular}}
\end{table}

%% file: Sections/4-method.tex
\section{Method}
\label{sec:method}
To answer \textbf{RQ2}, we propose the \textbf{R4R} framework.
We first provide an overview, then detail each step of the pipeline, and finally describe how to integrate R4R with existing GR methods.

\begin{algorithm}[t]
\caption{R4R inference}
\DontPrintSemicolon
\label{alg:r4r}
\KwIn{Query $q$; model $\mathcal{M}$; prompts $P_t,P_r,P_v,P_f$; constrained decoding strategy $\mathrm{cons}$; verify depth $t$ and round budget $T$}
\KwOut{Top-$k$ candidate $\mathrm{docids}$}
\rbtri*[l]{Think (Eq.\ref{eq:think})}
$(c_0,e_0) \leftarrow \mathcal{M}(P_t \,\|\, q)$ \;
\For{$i \leftarrow 1$ \KwTo $T$}{
  \rbtri*[l]{Retrieve (Eq.\ref{eq:retrieve})}
  $\mathrm{docid_i[1...k]} \leftarrow \mathcal{M}(P_r \,\|\, q \,\|\, c_{i-1};\mathrm{cons})$ \;
  \rbtri*[l]{Refine}
  $\hat{j} \leftarrow 0$ \;
  
  \For{$j \leftarrow 1$ \KwTo $t$}{
    \rtri*[l]{Verification}
    $r \leftarrow \mathcal{M}(P_v \,\|\, q \|\ \mathrm{docid_i[j]})$ \;
    \If{$r = \mathrm{irrelevant}$}{ 
        $\hat{j} \leftarrow j$\; 
        \textbf{break} \;
    }
    
  }
  \If{$\hat{j} =0$}{ \Return $\mathrm{docid_i[1{:}k]}$ \;}
  \rtri*[l]{Reflection (Eq.\ref{eq:reflection})}
  $(c_{i},e_{i}) \leftarrow \mathcal{M}(P_f \,\|\, q \,\|\, \mathrm{docid}_i[\hat{j}] \,\|\, c_{i-1} \,\|\, e_{i-1})$ \;
}
\Return $\mathrm{docid_{T}[1{:}k]}$
\end{algorithm}

\begin{figure*}[t]
    \centering
    \includegraphics[width=\textwidth]{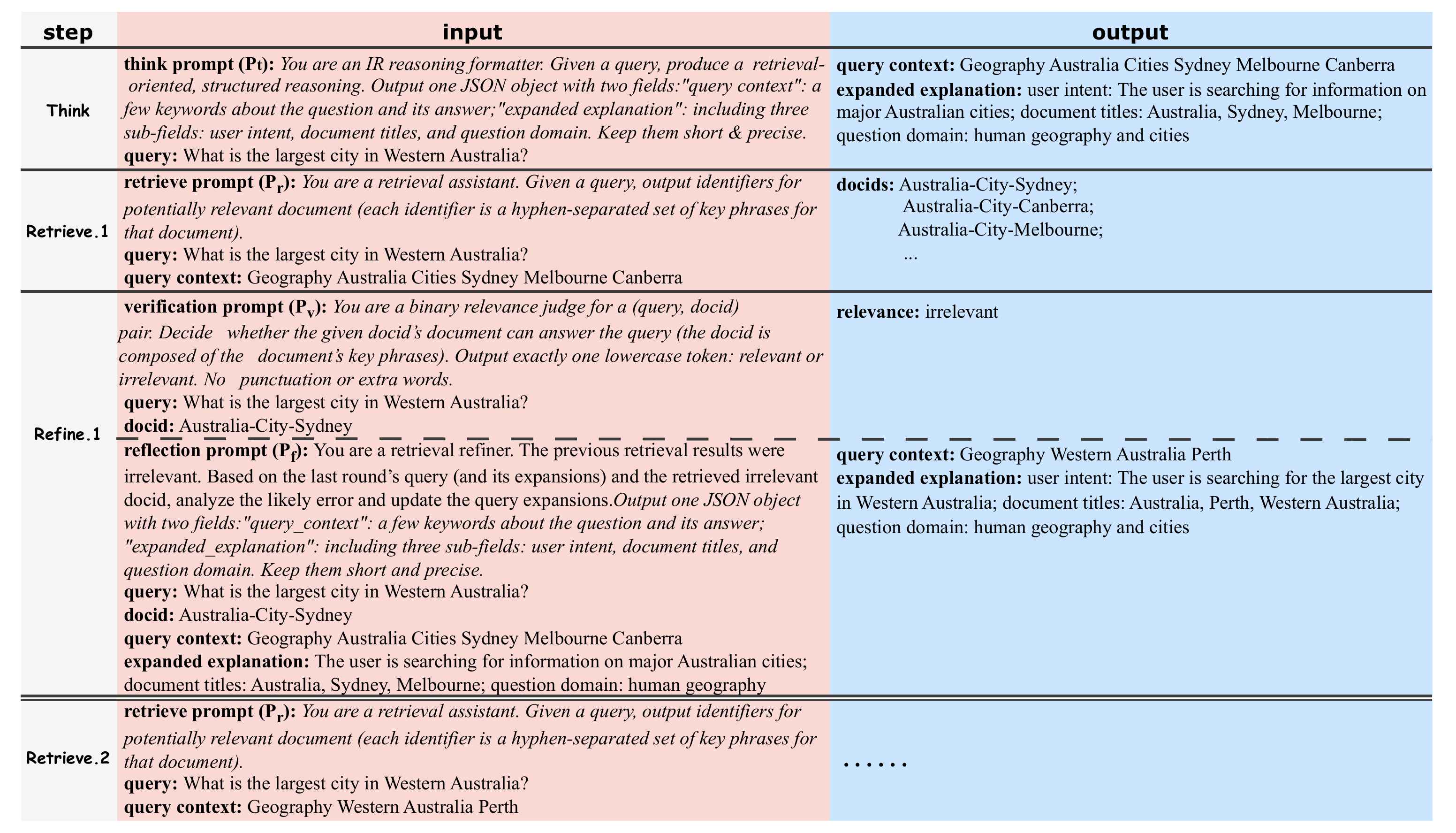}
    \caption{Illustration of the initial Think step and the Retrieve-Refine iteration in R4R.}
    \label{fig:overview}
\end{figure*}

\vspace{-2mm}
\subsection{Overview}
\label{subsec:overview}
R4R is built on an autoregressive LLM with reasoning abilities, integrated with a GR method that employs textual docids.  
During training, as described in Section~\ref{subsec:adapted_gr}, the standard GR training is adapted through instruction-tuning to grant the LLM retrieval capability, requiring no further training. 
At inference time, R4R uses the trained GR model to perform iterative retrieval under four specialized prompts: the thinking prompt $P_t$\footnote{This structured thinking prompt $P_t$ differs from the unconstrained decoding prompt $P_d$ (referred to as the one used in Direct CoT in Section~\ref{subsec:direct_cot}).}, the retrieval prompt $P_r$, the verification prompt $P_v$, and the reflection prompt $P_f$.

The overall workflow of R4R is shown in Figure~\ref{fig:overview} and Algorithm~\ref{alg:r4r}: 
Given a query, R4R first generates an initial query context and expanded explanation in the \emph{Think step}. 
It then enters an iterative loop: it performs constrained decoding conditioned on the original query and query context to retrieve candidate docids (\emph{Retrieve step}), after which the same model provides feedback on the docids to refine both the query context and expanded explanation (\emph{Refine step}). This loop continues until a stopping condition is met.

\vspace{-2mm}
\subsection{Think: initial reasoning generation}
R4R begins with a \emph{Think} step that formats and compresses intermediate reasoning into signals that are useful for retrieval and can be directly consumed by constrained decoding. 
This addresses the shortcomings of Direct CoT: as discussed in Section~\ref{subsec:direct_cot}, its verbose reasoning introduces intermediate steps that GR cannot effectively exploit and substantially increase decoding latency. 

Specifically, given a query $q$ and a thinking prompt $P_t$ , the model $\mathcal{M}$ performs a short unconstrained generation to produce a structured output: 
(i) a compact \textbf{query context} $c$ that briefly captures salient information the query likely points to and whose surface form is aligned with the docid, used as an auxiliary input for constrained decoding; and 
(ii) an \textbf{expanded explanation} $e$ that records key cues for interpreting the query and its context, later used for verification and reflection.
Formally,
\begin{equation}
    \langle c_0, e_0 \rangle \;=\; \mathcal{M}\!\left(P_t \, \Vert \, q\right),
\label{eq:think}
\end{equation}
where $c_0$ and $e_0$ denote the initial query context and expanded explanation, respectively. 
If structured parsing fails, we perform a minimal one-shot retry; if it still fails, we fall back to the raw query to ensure robustness. 

By formatting reasoning in this way, we decouple retrieval-critical signals from explanatory text: only the former ($c$)  is fed into constrained decoding for retrieval, while the latter ($e$) is reserved for the subsequent \emph{Refine} step. 
This reduces explanatory noise in the search process and improves overall efficiency.

\vspace{-2mm}
\subsection{Iterating Retrieve-Refine} 
After the \emph{Think} step, R4R enters an iterative loop. 
In iteration $i>0$, it uses the same model $\mathcal{M}$ as in \emph{Think} to perform the \emph{Retrieve} and \emph{Refine} steps, and repeats until a stopping condition is met.

\subsubsection{\textbf{Retrieve}}
In the \emph{Retrieve} step, R4R constructs the input by concatenating 
the retrieval prompt $P_r$, the original query $q$, and the current query context $c_{(i-1)}$. 
It then leverages existing constrained decoding methods to perform beam search, returning the top-$k$ candidate docids like in Eq.~\ref{eq:instruction_inference}. 
Formally,
\begin{equation}
    docid_i[1{:}k] = \mathrm{\mathcal{M}}\!\big(P_r \,\Vert\, q \,\Vert\, c_{i-1};\mathrm{cons}\big),
\label{eq:retrieve}
\end{equation}
where $docid_i[1{:}k]$ denotes the ranked list of $k$ candidates at iteration $i$. 
This step requires instruction-tuning for GR (as in Eq.~\ref{eq:instruction_training}). 
At inference time, the \emph{Retrieve} step follows the integrated GR method (including constrained decoding strategy), and the only change is that our input augments the raw query with the retrieval prompt $P_r$ and the query context $c_{(i-1)}$, rather than using the query alone.

\subsubsection{\textbf{Refine}}
After each \emph{Retrieve} step, R4R performs a \emph{Refine} step with two sub-steps: verification and reflection, using the same model $\mathcal{M}$.

\heading{Verification}
R4R conducts relevance \emph{verification} by scanning the top-$t$ candidates in order.\footnote{$t$ is typically smaller than the full candidate size $k$ because improving the query is most beneficial when errors appear at higher ranks, whereas acting on low-ranked errors may misdirect an otherwise correct retrieval.} 
The inputs are the original query $q$, a candidate docid and a verification prompt $P_v$, and the output is a binary relevance judgment $\{ \mathrm{relevant}, \mathrm{irrelevant} \}$. 
The scan stops immediately upon encountering the first irrelevant candidate. 
If all top-$t$ are relevant, the current retrieval results are returned and the loop terminates.

\heading{Reflection}
If an irrelevant candidate is found during verification, R4R performs \emph{reflection} to analyze the error and update both the query context and the expanded explanation. 
Specifically, given a reflection prompt $P_f$, the original query $q$, the first irrelevant docid $docid_{f}$, the current expanded explanation $e_i$ and the current query context $c_i$, the outputs are the updated query context $c_{i}$ and expanded explanation $e_{i}$:
\begin{equation}
    \langle c_{i},\, e_{i} \rangle \;=\; \mathcal{M}\!\big(P_f \,\Vert\, q \,\Vert\, docid_{f} \,\Vert\, c_{i-1} \Vert\, e_{i-1}\big).    
\label{eq:reflection}
\end{equation}
The update principle is to minimally edit only the key signals most related to the deviation, avoiding excessive changes that could cause drift. 
If structured parsing fails, we perform a minimal one-shot retry; if it still fails, the iteration terminates since carrying forward the previous $c_{i-1}$ and $e_{i-1}$ would merely repeat the same \emph{Retrieve} without benefit.

\subsubsection{\textbf{Iterating retrieve and refine}}
After updating the query context $c_i$ and the expanded explanation $e_i$ in round $i>0$, R4R proceeds to iteration $i{+}1$. 
The loop terminates when any of the following holds:
(i) in the \emph{Verification} sub-step, all top-$t$ candidates are judged relevant;
(ii) in the \emph{Reflection} sub-step, structured parsing fails; and
(iii) the round budget $T$ is reached.

The returned results of R4R in these cases are: 
\begin{equation}
\mathrm{Results} =
\begin{cases}
\mathrm{docid}_i[1{:}k], & \text{all relevant in round $i$;}\\[2pt]
\mathrm{docid}_i[1{:}k], & \text{parsing fails in round $i$;}\\[2pt]
\mathrm{docid}_T[1{:}k], & \text{round budget $T$ is reached.}
\end{cases}
\end{equation}

\subsection{Integration with existing GR methods}
R4R is designed for GR methods with textual docids. 
When integrating R4R, it needs to first train the model with the instruction-tuning objective in Eq.~\ref{eq:instruction_training}, and then reuses the existing method’s docid construction and constrained decoding strategy in the \emph{Retrieve} step. 
All other steps rely solely on the model’s inherent reasoning ability and require no additional training.

For GR with numeric docid, R4R is not directly applicable: such models typically lose general natural-language generation after training and can only produce numbers under unconstrained decoding, making \emph{Think} and \emph{Refine} infeasible. 
A possible workaround is to pair the GR with numeric docids with an external model: the two models respectively handle reasoning and constrained decoding, thereby executing an R4R-like iterative reasoning pipeline. 
However, this introduces extra models and system complexity (like multi-agent cooperation), which lies beyond the scope of this paper and is left for future work.

%% file: Sections/5-experimental_setup.tex
\section{Experimental setup}
\label{sec:experimental_setup}

\heading{Datasets} We first evaluate R4R on two widely-used public retrieval benchmarks: 
(i) \textbf{MS MARCO Passage \cite{msmarco}}, a large-scale web passage ranking dataset with real user queries and relevance annotations; and 
(ii) \textbf{Natural Questions (NQ) \cite{nq}}, which contains Google search queries paired with Wikipedia-derived answers and evidence passages. 
In addition, we assess R4R on a real-world \textbf{Taobao item-search benchmark} to evaluate practical performance, constructed from traffic in the digital-products vertical (2.5M query–item pairs).\footnote{Taobao is the largest C2C e-commerce platform in China. Further details regarding this benchmark are provided in Appendix \ref{app:taobao}.}
Unless otherwise noted, train/test splits follow the common practice for each dataset.

\input{tables/main_1}

\heading{GR baselines integrated with R4R}
Following Sections~\ref{sec:exploration} and~\ref{sec:method}, we reproduce four representative GR methods with textual docids (including their docid design and decoding constraint) and integrate them with R4R:
(i) \textbf{DSI \cite{dsi}}: a standard GR pipeline that trains an autoregressive LM to map both queries and documents to docids, and performs constrained decoding under prefix-trie at inference\footnote{Since the original DSI does not use textual docids, we adopt R4R’s default docid construction here to enable integration without introducing other confounding changes.};
(ii) \textbf{SEAL \cite{seal}}: shares the training setup with DSI and uses document titles as docids, and at inference, it replaces the trie constraint with an FM-index so decoding can resume from arbitrary positions, improving robustness to prefix errors;
(iii) \textbf{TSGen \cite{tsgen}}: shares the training setup with DSI, and at inference, it leverages an inverted index and term constraints to dynamically shrink the candidate space during generation, enabling constrained decoding from arbitrary offsets; and
(iv) \textbf{MINDER \cite{minder}}: similar to SEAL in training and inference, but assigns each document a set of multi-view textual docids and unifies cross-view ranking and merging.

\heading{External baselines for horizontal comparison}
For completeness, we also report representative methods from three IR paradigms as references:
(i) \textbf{Term-based}: \emph{BM25 \cite{bm25}} and \emph{DocT5Query \cite{doct5query}};
(ii) \textbf{Dense Retrieval}: \emph{DPR \cite{dpr}} and \emph{ANCE \cite{ance}}; and
(iii) \textbf{GR with numeric docids}: \emph{DSI (semantic identifiers)}, \emph{DSI-QG \cite{dsi-qg}}, \emph{LTRGR \cite{ltgtr}}, \emph{RIPOR \cite{ripor}} and \emph{PAG \cite{pag}}.
All reported numbers for these baselines are taken directly from published works \cite{ltgtr,novo,pag}.

\heading{Metrics} We report two standard retrieval metrics: 
(i) \textbf{Hits@$k$ \cite{dsi,seal}}: equals 1 if at least one relevant document appears in the top-$k$ candidates for a query, and 0 otherwise; we average over all queries; and
(ii) \textbf{MRR@$k$}: for each query, take the rank $r$ of the first relevant result within the top-$k$ and use $1/r$ (0 if no relevant result is found within $k$).

\heading{Implementation details} To fully leverage the reasoning capabilities of autoregressive models, we do not use the commonly adopted encoder–decoder architectures in GR~\cite{dsi,seal,nci,minder,ripor}. 
Instead, we choose larger decoder-only LMs as the backbone for R4R. 
Specifically, we use Llama-3.1-8B \cite{llama3} as the backbone on NQ and MS~MARCO. 
For the Taobao item-search dataset (Chinese), we employ Qwen3-14B \cite{qwen3} as the backbone.

For GR training, we adopt a standard LoRA fine-tuning setup \cite{lora}. 
The rest of the training configuration follows each GR method’s original recipe, with the only modification being the addition of an explicit retrieval instruction in the input to enable instruction-tuning (see Section~\ref{sec:exploration}). 
For inference, in \emph{Think} and \emph{Refine}, we use greedy decoding. 
In \emph{Retrieve}, we perform constrained beam search (the exact constraint structure follows the integrated GR method) with beam size $k{=}20$ and a default verify depth of $t{=}3$. 
We set the default maximum number of R4R iterations to $T{=}3$. 
Experiments on NQ and MS~MARCO are run on 2$\times$NVIDIA A100 and experiments on the Taobao dataset are run on 2$\times$NVIDIA H20.

%% file: tables/main_1.tex
\begin{table*}[t]
\centering
\caption{Performance of representative GR methods before and after integrating R4R on NQ and MS MARCO.}
\label{tab:main_results_1}
\setlength{\tabcolsep}{12pt}
  \renewcommand{\arraystretch}{1.1}
\begin{tabular}{lccccccc}
\toprule
\multirow{2}{*}{\textbf{Method}} & \multicolumn{4}{c}{\textbf{NQ}} & \multicolumn{3}{c}{\textbf{MS~MARCO}}\\
\cmidrule(lr){2-5}\cmidrule(l){6-8}
 & Hits@1 & Hits@5 & Hits@20 & MRR@10 & Hits@5 & Hits@10 & MRR@10 \\
\midrule
DSI-text            & 46.0 & 59.6 & 75.3 & 56.3 & 35.9 & 55.8 & 34.1 \\
\quad + R4R                   & 46.6 & 59.6 & 76.9 & 58.1 & 37.2 & 55.8 & 35.2 \\
\hline
SEAL                & 50.9 & 63.5 & 79.3 & 61.2 & 41.4 & 61.1 & 37.2 \\
\quad + R4R                   & \underline{53.1} & 66.0 & \underline{81.2} & 65.3 & \underline{44.3} & 63.9 & \textbf{38.5} \\
\hline
MINDER              & 50.0 & 66.8 & 80.0 & 62.5 & \underline{44.3} & \underline{64.7} & 37.9 \\
\quad + R4R                   & \textbf{53.8} & \textbf{69.3} & 80.8 & \underline{67.7} & \textbf{45.7} & 64.1 & \underline{38.1} \\
\hline
TSGen             & 48.8 & 67.1 & 79.7 & 64.6 & 42.2 & 64.0 & 35.1 \\
\quad + R4R                   & 52.3 & \underline{69.1} & \textbf{81.6} & \textbf{68.5} & 44.2 & \textbf{66.7} & 36.3 \\

\bottomrule
\end{tabular}
\end{table*}

%% file: Sections/6-experiments.tex
\input{tables/reference}
\section{Experimental results}
\label{sec:experiments}

In this section, we conduct experiments to verify the effectiveness of R4R on NQ, MS~MARCO and Taobao item-search dataset.

\vspace{-2mm}
\subsection{Baseline comparison}

\heading{Results on NQ and MS~MARCO} We integrate R4R with four GR methods with textual docids (DSI-text, SEAL, MINDER, TSGen). 
Table~\ref{tab:main_results_1} shows: 
(i) R4R delivers stable improvements over all four GR methods across both datasets. For example, on NQ, SEAL+R4R improves Hits@1 and MRR@10 by 2.2\% and 4.1\%, respectively. On MS~MARCO, TSGen+R4R improves Hits@5 and MRR@10 by 2.0\% and 1.2\%, respectively;
(ii) R4R yields a slightly larger average improvement for GR on NQ than on MS MARCO. This may be because NQ queries are natural questions that typically require inferring the answer before locating supporting documents, and R4R’s reasoning process helps bridge this gap; and
(iii) when the constrained decoding doesn't force a fixed order (like in SEAL, MINDER, and TSGen), R4R usually helps more than it does with DSI, which does require a fixed order. This indicates that R4R’s compact, docid-aligned query context is particularly effective when multiple equivalent generation orders are allowed.

We compile reported results of representative methods from other retrieval paradigms on NQ and MSMARCO \cite{ltgtr,novo,pag} and compare them against two GR methods enhanced with R4R.
As shown in Table~\ref{tab:reference}, we can observe that compared to representative methods from other paradigms, GR + R4R achieves strong overall quality. For instance, on MS~MARCO, the best integration (SEAL+R4R) attains MRR@10 \(= 38.5\), matching the strongest baseline.

\input{tables/main_2}

\heading{Results on Taobao item-search dataset} To assess whether R4R applies beyond standard QA-style retrieval, we further evaluate it on the Taobao item-search dataset.
From Table~\ref{tab:main_results_2}, we observe the same trend as on the public benchmarks: R4R improves all four GR methods with textual docids on Taobao, for example, SEAL+R4R improves Hits@1 and MRR@10 by 2.5\% and 2.7\%, respectively. 
Moreover, GR+R4R clearly outperforms baselines from other retrieval paradigms on this dataset, showing strong effectiveness for item search. 
This suggests that R4R is not only effective on public benchmarks but also practical for other real-world scenarios.

\begin{figure*}[t]
  \centering
  \begin{subfigure}{0.47\textwidth}
    \includegraphics[width=\linewidth]{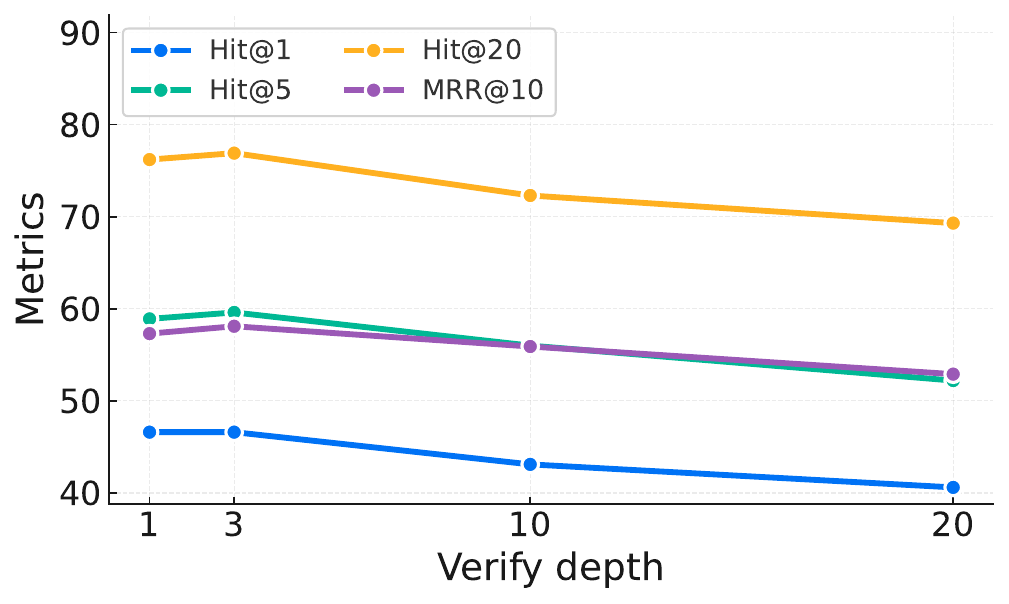}
  \end{subfigure}
  \begin{subfigure}{0.47\textwidth}
    \includegraphics[width=\linewidth]{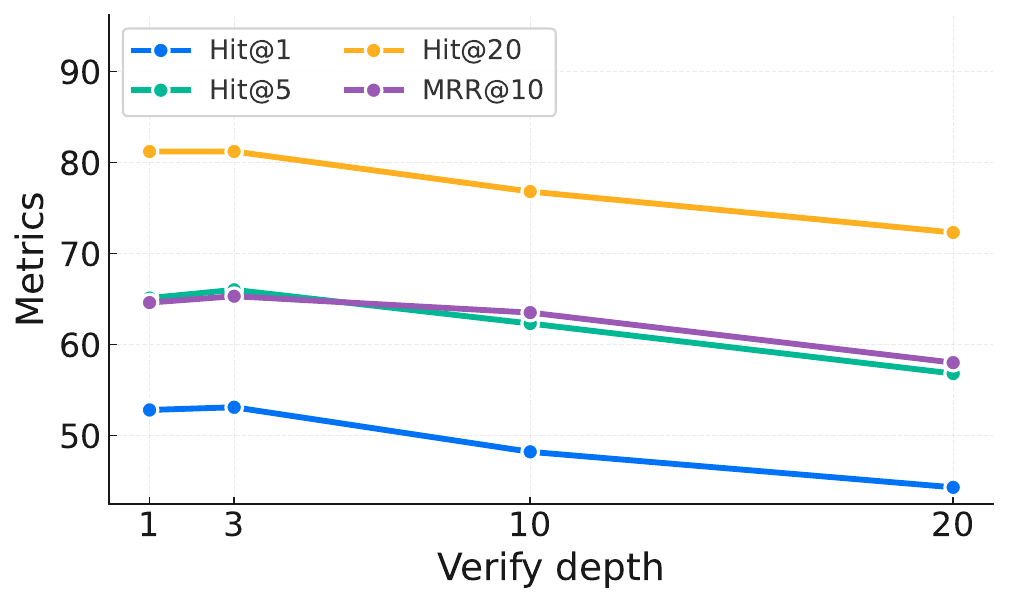}
  \end{subfigure}
  \caption{Performance trends of R4R as the verify depth varies on NQ. The left plot shows R4R integrating DSI and the right plot shows R4R integrating SEAL.}
  \label{fig:verify}
\end{figure*}

\vspace{-2mm}
\subsection{Ablation study}

We perform an ablation study by sequentially removing individual R4R components that can be omitted without disrupting the pipeline on NQ. 
(i) \emph{w/o query context}: generate only the expanded explanation in \emph{Think} and \emph{Refine}, and use it for both \emph{Retrieve} and \emph{Refine}; 
(ii) \emph{w/o expanded explanation}: generate only the query context in \emph{Think} and \emph{Refine}, and use it for both \emph{Retrieve} and \emph{Refine}; and 
(iii) \emph{w/o verification}: skip relevance judgment in \emph{Refine} and directly perform reflection on all candidates. 

\input{tables/ablation}

As shown in Table~\ref{tab:ablation}, we observe: 
(i) the full R4R with all components consistently performs best, demonstrating the necessity of each component; 
(ii) using only the expanded explanation causes a sharp drop (even below the original methods), because it  carries support signals not tailored for retrieval and thus is unsuitable as auxiliary input to \emph{Retrieve}; 
(iii) conversely, using only the query context leads to a smaller yet noticeable decrease, likely because \emph{Refine} lacks the supporting cues needed for effective updates; and 
(iv) skipping verification also yields a significant degradation, as the model fails to both judge candidate quality and update the cues reliably in a single step.

\vspace{-2mm}
\subsection{Impact of different factors}

\heading{Impact of verify depth}
We vary the depth \(t \in \{1,3,10,20\}\) while keeping the rest of the pipeline unchanged.
Results on NQ are shown in Figure \ref{fig:verify}: 
(i) with shallow verification ( \(t=1\) or \(3\) ), R4R brings modest improvements; 
(ii) with larger depths, the performance of R4R degrades noticeably; and 
(iii) in the first half of the \(\mathrm{Hits@1}\) curve is nearly flat, whereas in the second half \(\mathrm{Hits@1}\) drops the fastest. 
We attribute this to the following: when the first candidate is already relevant (i.e., the first irrelevant candidate appears at a lower rank), updating the query context based on that lower-ranked negative may mislead the subsequent \emph{Retrieve}.
Therefore, we adopt \(t=3\) as the default trade-off setting.

\input{tables/termination}

\heading{Iteration termination analysis}
We report the proportions of three termination cases in the iterative process on NQ:
(i) all candidates being judged relevant; (ii)  the round budget being exhausted; and (iii) a parsing failure. 
From Table~\ref{tab:termination}, we observe: 
(i) the all-relevant case dominates and is far higher than the round-budget case, indicating that our round budget is reasonably generous; and 
(ii) the parse-failure ratio is the lowest, suggesting that the structured outputs together with the minimal retry mechanism are stable and reliable in practice.

\input{tables/training}
\input{tables/rewriting}
\heading{Impact of instruction-tuning}
Finally, we include a supplementary study on the impact of instruction-tuning on NQ.
Specifically, for each of the four GR methods, we compare two fine-tuning approaches (standard training and instruction-training) and report performance without R4R to isolate the effect of the training scheme. 
Results are shown in Table~\ref{tab:training_effect}.
We observe that the differences induced by switching the training approaches are very small. 
For some models, instruction-style training even yields slight gains (albeit not statistically significant).
This indicates that adopting instruction training to preserve the model’s reasoning ability does not degrade the base retrieval performance.

\heading{Comparison to other query rewriting methods}
R4R can be viewed as a multi-round query rewriting approach. 
We therefore compare it against 
(i) \emph{no rewriting}: directly using the raw query (single round only); 
(ii) \emph{pseudo-relevance feedback (PRF)}: performing one retrieval to obtain pseudo-relevant documents and applying sparse expansion (single round only); and 
(iii) \emph{external LLM rewriting}: invoking an external LLM API to rewrite the query, following the same workflow as R4R except that the \emph{Think} and \emph{Refine} steps use an external LLM. 
Results on NQ are shown in Table~\ref{tab:rewrite}: 
(i) in the single-round setting, the two generation-based methods (R4R and external LLM rewriting) outperform the traditional rewriting baseline (PRF); and 
(ii) external LLM rewriting yields slightly higher accuracy than R4R, but R4R relies only on the base LLM with no external calls, achieving nearly comparable gains at a better cost-effectiveness trade-off.

\begin{figure}[h]
    \centering
    \includegraphics[width=0.96\linewidth]{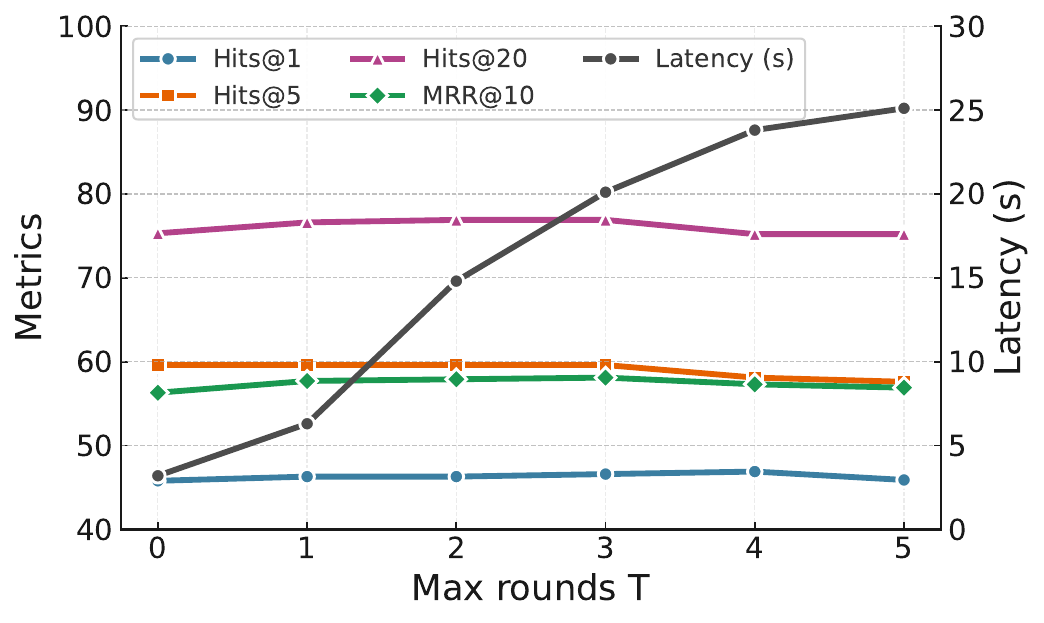}
    \caption{Performance and efficiency trends of DSI+R4R as the round budget  varies on NQ.}
    \label{fig:rounds}
\end{figure}

\heading{Impact of iteration rounds on effectiveness and efficiency}
We study how the maximum number of rounds affects R4R while accounting for efficiency on NQ. 
Specifically, we set \(T\!\in\!\{0,1,2,3,4,5\}\), and report both retrieval metrics and decoding latency. 
Figure~\ref{fig:rounds} shows: 
(i) retrieval metrics rise with \(T\) at first, but peak around \(T{=}3\)–\(4\) and then begin to decline; and
(ii) latency does not increase strictly linearly (since some iterations terminate early), yet compared with the baseline, too many rounds still introduce nearly unacceptable overhead.
Therefore, we set the default maximum number of rounds to \(T{=}3\).

%% file: tables/reference.tex
\begin{table}[b]
\centering
\caption{Performance comparison of GR+R4R with classic IR methods for reference (values for classic methods are taken from the published works \cite{ltgtr,novo,pag}).}
\label{tab:reference}
\setlength{\tabcolsep}{6pt}
\begin{tabular}{lcccc}
\toprule
\multirow{2}{*}{\textbf{Method}} & \multicolumn{2}{c}{\textbf{NQ}} & \multicolumn{2}{c}{\textbf{MS~MARCO}}\\
\cmidrule(lr){2-3}\cmidrule(l){4-5}
 & Hits@5 & Hits@20 & Hits@10 & MRR@10 \\
\midrule\midrule
\multicolumn{5}{l}{\textit{term-based retrieval}}\\
BM25                          & 43.6 & 62.9 & 69.1 & 18.5 \\
DocT5Query                    & 50.7 & 68.6 & \underline{75.1} & 27.2 \\
\midrule\midrule
\multicolumn{5}{l}{\textit{dense retrieval}}\\
DPR                           & 68.3 & 80.1 & 63.3 & 31.7 \\
ANCE                          & \textbf{69.2} & 80.1 & \textbf{75.7} & 33.0 \\
\midrule\midrule
\multicolumn{5}{l}{\textit{generative retrieval}}\\
DSI-semantic                  & 28.3 & 47.3 & 43.6 &  --  \\
DSI-QG                        & 35.5 & 52.7 &  --  & 10.5 \\
LTRGR                         & 68.8 & 80.3 &  --  & 25.5 \\
RIPOR                         &  --  &  --  & 56.2 & 33.3 \\
PAG                           &  --  &  --  & 67.0 & \textbf{38.5} \\
\midrule\midrule
\multicolumn{5}{l}{\textit{R4R}}\\
SEAL+R4R                      & 66.0 & \underline{81.2} & 63.9 & \textbf{38.5}  \\
TSGen+R4R                     & \underline{69.1} & \textbf{81.6} & 66.7 & \underline{36.3}  \\
\bottomrule
\end{tabular}
\vspace{-1mm}
\end{table}

%% file: tables/main_2.tex
\begin{table}[h]
\centering
\caption{Performance comparison of GR+R4R and classic IR methods on Taobao item search.}
\label{tab:main_results_2}
\setlength{\tabcolsep}{6pt}
\begin{tabular}{lcccc}
\toprule
\textbf{Method} & Hits@1 & Hits@5 & Hits@20 & MRR@10 \\
\midrule
DSI-text            & 27.3 & 49.3 & 65.8 & 28.2 \\
\quad + R4R         & 28.1 & 49.6 & 66.3 & 30.5 \\
\hline
SEAL                & 30.0 & 55.8 & 69.3 & 31.0 \\
\quad + R4R         & 32.5 & 57.0 & 71.5 & 33.7 \\
\hline
MINDER              & 32.6 & 56.1 & 69.5 & 36.5 \\
\quad + R4R         & \underline{33.7} & \underline{59.6} & \textbf{73.1} & 38.1 \\
\hline
TSGen             & 33.1 & 57.1 & 71.0 & 37.6 \\
\quad + R4R         & \textbf{34.2} & 59.2 & \underline{72.5} & \underline{39.3} \\
\midrule\midrule
BM25                & 18.9 & 42.8 & 55.1 & 28.3 \\
DPR                 & 29.1 & \textbf{62.8} & \textbf{73.1} & \textbf{39.5} \\
DSI-semantic        & 25.7 & 43.6 & 58.3 & 31.0 \\
DSI-QG              & 24.6 & 41.3 & 50.7 & 29.8 \\
\bottomrule
\end{tabular}
\vspace{-2mm}
\end{table}

%% file: tables/ablation.tex
\begin{table}[h]
\centering
\caption{Results of ablation experiments of R4R on NQ.}
\label{tab:ablation}
\resizebox{\columnwidth}{!}{
\begin{tabular}{lcccc}
\toprule
\textbf{Method} & Hits@1 & Hits@5 & Hits@20 & MRR@10 \\
\midrule
\multicolumn{5}{l}{\textbf{DSI-text}}\\
+R4R                    & \textbf{46.6} & \textbf{59.6} & \textbf{76.9} & \textbf{58.1} \\
w/o query context       & 23.6 & 28.7 & 36.1 & 30.0 \\
w/o expanded explanation& 46.0 & 59.6 & \textbf{76.9} & 57.7 \\
w/o verification        & 35.6 & 40.5 & 65.1 & 41.2 \\
\addlinespace[2pt]
\hline
\multicolumn{5}{l}{\textbf{SEAL}}\\
+R4R                    & \textbf{53.1} & \textbf{66.0} & \textbf{81.2} & \textbf{65.3} \\
w/o query context       & 27.3 & 33.5 & 45.2 & 33.6 \\
w/o expanded explanation& 51.3 & 64.2 & 79.6 & 63.7 \\
w/o verification        & 32.6 & 41.3 & 68.2 & 41.3 \\
\addlinespace[2pt]
\hline
\multicolumn{5}{l}{\textbf{MINDER}}\\
+R4R                    & \textbf{53.8} & \textbf{69.3} & \textbf{80.8} & \textbf{67.7} \\
w/o query context       & 24.8 & 31.4 & 42.9 & 32.5 \\
w/o expanded explanation& 53.5 & 68.3 & 79.2 & 66.0 \\
w/o verification        & 36.8 & 44.0 & 71.2 & 46.6 \\
\addlinespace[2pt]
\hline
\multicolumn{5}{l}{\textbf{TSGen}}\\
+R4R                    & 52.3 & \textbf{69.1} & \textbf{81.6} & \textbf{68.5} \\
w/o query context       & 26.2 & 34.8 & 48.0 & 32.6 \\
w/o expanded explanation& \textbf{52.5} & 68.4 & 80.8 & 67.2 \\
w/o verification        & 22.3 & 30.5 & 46.2 & 29.8 \\
\bottomrule
\end{tabular}}
\end{table}

%% file: tables/termination.tex
\begin{table}[t]
\centering
\caption{Termination conditions for R4R iterations on NQ.}
\label{tab:termination}
\resizebox{\columnwidth}{!}{
\begin{tabular}{lccc}
\toprule
\textbf{Method} & All relevant & Round budget exceeded & Parse failure \\
\midrule
DSI-text + R4R & \textbf{72.8} & 23.5 & 3.7 \\
SEAL + R4R     & \textbf{80.2} & 14.6 & 5.2 \\
MINDER + R4R   & \textbf{78.8} & 16.5 & 4.7 \\
TSGen + R4R    & \textbf{79.1} & 17.2 & 3.7 \\
\bottomrule
\end{tabular}}
\vspace{-2mm}
\end{table}

%% file: tables/training.tex
\begin{table}[h]
\centering
\caption{Effect of instruction-tuning on NQ.}
\label{tab:training_effect}
\resizebox{\columnwidth}{!}{
\begin{tabular}{l l cccc}
\toprule
\textbf{Method} & Training & Hit@1 & Hit@5 & Hit@20 & MRR@10 \\
\midrule
\multirow{2}{*}{DSI-text}
& Standard      & 45.8 & 59.6 & 75.3 & 56.3 \\
& Instruction   & 46.0 & 59.6 & 75.3 & 56.3 \\
\addlinespace[2pt]
\multirow{2}{*}{SEAL}
& Standard      & 51.2 & 61.8 & 78.7 & 61.0 \\
& Instruction   & 50.9 & 63.5 & 79.3 & 61.2 \\
\addlinespace[2pt]
\multirow{2}{*}{MINDER}
& Standard      & 48.2 & 66.9 & 79.3 & 62.1 \\
& Instruction   & 50.0 & 66.8 & 80.0 & 62.5 \\
\addlinespace[2pt]
\multirow{2}{*}{TSGen}
& Standard      & 48.6 & 66.7 & 79.5 & 63.1 \\
& Instruction   & 48.8 & 67.1 & 79.7 & 64.6 \\
\bottomrule
\end{tabular}}
\vspace{-2mm}
\end{table}

%% file: tables/rewriting.tex
\begin{table}[t]
\centering
\caption{Comparison of R4R with classic query rewriting
methods on NQ.}
\label{tab:rewrite}
\resizebox{\columnwidth}{!}{
\begin{tabular}{lcccc}
\toprule
\textbf{Single round} & Hits@1 & Hits@5 & Hits@20 & MRR@10 \\
\midrule
No rewrite     & 45.8 & \textbf{59.6} & 75.3 & 56.3 \\
Sparse rewrite & 43.7 & \underline{55.2} & 70.7 & 51.5 \\
LLM rewrite    & \textbf{47.6} & \textbf{59.6} & \underline{76.3} & \textbf{58.1} \\
R4R            & \underline{46.3} & \textbf{59.6} & \textbf{76.6} & \underline{57.7} \\
\midrule
\midrule
\textbf{Iterative} & Hits@1 & Hits@5 & Hits@20 & MRR@10 \\
\midrule
LLM rewrite    & \textbf{47.2} & \textbf{61.0} & \textbf{77.2} & \textbf{59.3} \\
R4R            & \underline{46.6} & \underline{59.6} & \underline{76.9} & \underline{58.1} \\
\bottomrule
\end{tabular}}
\end{table}

%% file: Sections/2-related_works.tex
\vspace{-2mm}
\section{Related works}
\vspace{-2mm}
\label{sec:related}

\heading{Generative retrieval}
\label{subsec:gr}
GR uses an autoregressive LLM to encode the corpus into model parameters during training and, at inference, generates document identifiers (docids) end-to-end given a query \cite{dsi,seal,nci}. Some works focus on the docid design of GR: 
(i) methods that use \emph{numeric docids \cite{dsi,nci,ltgtr,listwise-gr}}, assigning each document an integer or employing structured codebooks derived from hierarchical k-means, product quantization (PQ), or residual quantization (RQ). 
These approaches often weaken the model’s general text generation ability after training; and 
(ii) methods that use \emph{textual docids \cite{seal,tsgen,minder,corpus-brain}}, such as titles, n-grams, or URLs. By retaining document priors in textual form, these methods better align with the LLM’s pretraining objective and thus partially preserve the model’s generative capability. 
Other works focus on improving constrained decoding of GR. 
For example, 
(i) SEAL \cite{seal} replaces the prefix trie with an FM-index, enabling decoding from arbitrary positions in the docid; 
(ii) TSGen \cite{tsgen} introduces an inverted index to likewise support arbitrary-position decoding; and 
(iii) Minder \cite{minder} assigns multiple views of docids to each document and unifies ranking across these views. 
In this work, we focus on using CoT to improve GR’s retrieval performance, so we scope our study to GR with textual docids. 
In addition, different constrained decoding methods are largely compatible with our R4R.

\heading{Reasoning enhancements}
\label{subsec:reasoning} Chain-of-Thought (CoT) \cite{cot} refers to explicitly generating step-by-step intermediate reasoning before the final answer. 
CorpusLM \cite{corpuslm} can be viewed as an early attempt to combine CoT with GR by appending reasoning in the downstream QA stage after GR returns candidates. 
To our knowledge, there is no work that leverages CoT directly within the retrieval stage to improve retrieval quality and this paper fills this gap. 
Self-refine and self-verification methods aim to iteratively improve model outputs under no or weak supervision via self-assessment and small edits \cite{self-refine,chain-of-verification,reflexion}. 
In GR, self-retrieval \cite{self-retrieval} introduces self-evaluation but stops at post-retrieval quality judgment without feeding the decision back into the retrieval signal. 
In contrast, we discuss how to act on relevance judgments to improve retrieval.

%% file: Sections/7-conclusion.tex
\vspace{-2mm}
\section{Conclusion}
\label{sec:conclusion}
We studied whether explicit reasoning can improve GR and how to make it accurate and efficient in practice. 
We present R4R, a reasoning-augmented framework for GR that allows a single LLM to iteratively generate relevant docids and refine their own outputs. 
We demonstrate the ease of use of R4R across NQ, MS MARCO, and a real-world Taobao item-retrieval dataset, yielding consistent retrieval gains with minimal additional decoding overhead.

Our study has several limitations that suggest directions for future work: 
\begin{enumerate*}[label=(\roman*)]
    \item we focus on GR with textual docids, GR with numeric docids is included only as a reference baseline, and designing effective reasoning for such identifiers remains an open challenge; 
    \item R4R currently assumes relatively large, reasoning-capable LLMs, which may incur prohibitive costs in some industrial settings; and
    \item updating the retrieval prompt bundle based solely on the first erroneous candidate is imperfect, when the first irrelevant item occurs at a low-ranked position while higher-ranked results are already correct, this signal can misdirect the process. Developing more robust refinement strategies is a worthwhile challenge.
\end{enumerate*}

%% file: Sections/appendix.tex
\newpage
\appendix

\section{Prompt templates in our exploration}
\label{app:prompt}
In this section, we provide the prompts used in Section~\ref{sec:exploration}: the retrieval prompt $P_r$ (identical to the inference-time retrieval prompt) and the indexing prompt $P_i$, as well as the Direct CoT prompt $P_d$.
\lstset{basicstyle=\ttfamily\small,frame=single,aboveskip=4pt,belowskip=4pt}
\begin{lstlisting}[caption=Retrieval Prompt $P_r$,label=lst:prompt_pr]
You are a retrieval assistant. 
Given a query, output identifiers for potentially
relevant document (each identifier is a hyphen-
separated set of key phrases for that document).
\end{lstlisting}

\lstset{basicstyle=\ttfamily\small,frame=single,aboveskip=4pt,belowskip=4pt}
\begin{lstlisting}[caption=Indexing Prompt $P_i$,label=lst:prompt_pi]
You are a retrieval assistant. 
Given a document, output identifiers for 
potentially relevant document (each identifier is 
a hyphen-separated set of key phrases for that 
document).
\end{lstlisting}

\lstset{basicstyle=\ttfamily\small,frame=single,aboveskip=4pt,belowskip=4pt}
\begin{lstlisting}[caption=Direct CoT Prompt $P_d$,label=lst:prompt_pd]
You are a QA assistant. 
Given a query, think step by step about the answer 
and which documents are likely to contain it.
\end{lstlisting}

\section{Taobao item-search benchmark}
\label{app:taobao}
Taobao is one of the largest e-commerce platforms in China. 
On Taobao, item search is the primary channel connecting users with items of interest. 
Based on real-world Taobao data, we introduce an item-search dataset and a corresponding item-retrieval task.

The dataset is constructed from real user search logs that record the user’s query text, clicked items, and purchased items. 
We define the latter two together as the relevant items. 
In other words, each record contains two fields: 
(i) \emph{query}: the user’s actual query text; and 
(ii) \emph{relevant item list}: an unordered list comprising the items clicked or purchased for that query. 
To reduce deployment complexity and enable rapid validation, we limit experiments to consumer electronics, totaling 2.5 million query-item pairs.

The item-retrieval task is defined as identifying the most relevant item categories for subsequent ranking, given a user query (as required by Taobao’s real business needs). 
To this end, we first perform hierarchical clustering over items in the dataset to obtain a taxonomy, then replace each query-item pair with a query-cluster pair. 
We subsequently train and evaluate on these query-cluster pairs.